\documentclass[twocolumn]{aastex631}

\usepackage{multirow} 
\usepackage{makecell}
\usepackage{graphicx}

\newcommand{\MyPath}{./}
\graphicspath{{./}}

\newcommand{\SiXIV}{{\fontfamily{qcr}\selectfont Si XIV}}   
\newcommand{\NeX}{{\fontfamily{qcr}\selectfont Ne X}}       
\newcommand{\SXVI}{{\fontfamily{qcr}\selectfont S XVI}}     

\newcommand{\CIAO}{{\fontfamily{qcr}\selectfont CIAO}}          
\newcommand{\CALDB}{{\fontfamily{qcr}\selectfont CALDB}}        
\newcommand{\XSPEC}{{\fontfamily{qcr}\selectfont XSPEC}}        

\newcommand{\steppar}{{\fontfamily{qcr}\selectfont steppar}}    
\newcommand{\pow}{{\fontfamily{qcr}\selectfont pow}}            

\newcommand{\Ka}{{\fontfamily{qcr}\selectfont K$\alpha$}}   

\newcommand{\Chandra}{{\fontfamily{qcr}\selectfont Chandra}}    
\newcommand{\XMMN}{{\fontfamily{qcr}\selectfont XMM-Newton}}    

\newcommand{\MEG}{{\fontfamily{qcr}\selectfont MEG}}    
\newcommand{\LETG}{{\fontfamily{qcr}\selectfont LETG}}  

\newcommand{\ACISSHETGMEG}{{\fontfamily{qcr}\selectfont ACIS-S HETG-MEG}}   
\newcommand{\ACISSHETG}{{\fontfamily{qcr}\selectfont ACIS-S HETG}}   
\newcommand{\ACISSLETG}{{\fontfamily{qcr}\selectfont ACIS-S LETG}}          
\newcommand{\ACISS}{{\fontfamily{qcr}\selectfont ACIS-S}}                   

\newcommand{\warm}{{\fontfamily{qcr}\selectfont warm}}                          
\newcommand{\warmhot}{{\fontfamily{qcr}\selectfont warm-hot}}                   
\newcommand{\hot}{{\fontfamily{qcr}\selectfont hot}}                            

\newcommand{\combinegratingspectra}{{\fontfamily{qcr}\selectfont combine$\_$grating$\_$spectra}}    
\newcommand{\chandrarepro}{{\fontfamily{qcr}\selectfont chandra$\_$repro}}                          

\newcommand{\A}{$\mathrm{\AA}$}                     

\begin{document}

%
%

\title{Detection of a Super-Virial Hot Component in the Milky Way Circumgalactic Medium Along Multiple Sight-Lines by Using the Stacking Technique}

\shorttitle{Super-Virial Hot Component in the Milky Way CGM}

%
%

\shortauthors{Lara-DI et al.}

\author[0000-0001-6995-2366]{Armando Lara-DI}
\affiliation{Instituto de Astronomia, Universidad Nacional Autonoma de México, 04510 Mexico City, Mexico}

\author[0000-0002-4822-3559]{Smita Mathur}
\affiliation{Astronomy Department, The Ohio State University, Columbus, OH 43210}

\author[0000-0001-6291-5239]{Yair Krongold}
\affiliation{Instituto de Astronomia, Universidad Nacional Autonoma de México, 04510 Mexico City, Mexico}

\author[0000-0002-9069-7061]{Sanskriti Das}
\affiliation{Astronomy Department, The Ohio State University, Columbus, OH 43210}
\affiliation{Kavli Institute for Particle Astrophysics \& Cosmology, Stanford University, 452 Lomita Mall, Stanford, CA 94305, USA}

\author[0000-0003-1880-1474]{Anjali Gupta}
\affiliation{Astronomy Department, The Ohio State University, Columbus, OH 43210}

%
%

\begin{abstract}
    
    The study of the elusive \hot\ component ($T \gtrsim 10^7$ K) of the Milky Way circumgalactic medium (CGM) is a novel topic to understand Galactic formation and evolution. In this work, we use the stacking technique through 46 lines of sight with \Chandra\ \ACISSHETG\ totaling over 10Ms of exposure time and 9 lines of sight with \ACISSLETG\ observations totaling over 1Ms of exposure time, to study in absorption the presence of highly ionized metals arising from the super-virial temperature phase of the CGM. Focusing  in the spectral range  $4 - 8$ \AA, we were able to confirm the presence of this \hot\ phase with high significance. We detected transitions of \SiXIV\ \Ka\ (with total significance of 6.0$\sigma$) and, for the first time, \SXVI\ \Ka\ (total significance 4.8$\sigma$) in the rest frame of our own Galaxy. For \SXVI\ \Ka\ we found a column density of $1.50^{+0.44}_{-0.38} \times 10^{16} \mathrm{cm}^{-2}$. For \SiXIV\ \Ka\ we measured a column density of $0.87\pm{0.16} \times 10^{16} \mathrm{cm}^{-2}$. The lines of sight used in this work are spread across the sky, probing widely separated regions of the CGM. Therefore, our results indicate that this newly discovered \hot\ medium extends throughout the halo, and is not related only to the Galactic Bubbles. The hot gas location, distribution, and covering factor, however, remain unknown. This component might contribute significantly to the missing baryons and metals in the Milky Way.
    
\end{abstract}

%
%


%
%

\section{Introduction \label{sec:intro}}

    The circumgalactic medium (CGM) is a massive, diffuse halo composed of several gas phases with different temperatures and ionization states. It is defined as the gas outside the galactic disc and inside the virial radius of the Galaxy (\citealp{Tumlinson2017}). 

    The study of the CGM provides insights into galaxies' physical and chemical history. Numerical simulations (e.g., \citealp{Stinson2012}) suggest that the CGM is constantly perturbed and enriched with gas and dust from feedback processes from starburst or AGN processes of the Galactic disk. This enrichment makes the CGM an important clue to understand star formation history and galactic evolution theories (e.g., \citealp{Stinson2012}; \citealp{Fabrega2016}; \citealp{Tumlinson2017}). 

    The inflows and outflows of gas make the CGM a multiphase medium. Theoretical models predict that the CGM is composed of two phases of collisionally ionized gas; the first one is a \warm\ phase at temperature ($T$) $\sim 10^5$ K, and the second one is the \warmhot\ phase with temperatures around $10^6$ K, close to the virial temperature (\citealp{Stinson2012}; \citealp{Fabrega2016}). These two phases have been observed as $z=0$ absorption lines in the UV and X-ray spectra of distant quasars (e.g. \citealp{Savage2003}; \citealp{Gupta2012}; \citealp{Nicastro2016}; \citealp{Tumlinson2017}; \citealp{Das2019}).

    Because of its importance for galaxy evolution, the CGM of galaxies has been the centre of attention in recent studies. These studies suggest that it is an extensive reservoir of gas and dust where the missing baryons and missing metals could reside (\citealp{Tumlinson2017}; \citealp{Das2019}). 
    
    The missing baryons problem is one of the central puzzles in the frame of cosmic and galactic evolution. The number of baryons predicted by Nucleosynthesis of the Big Bang Theory, corresponds to that inferred from the density fluctuations of the Cosmic Microwave Background (CMB) (e.g., \citealp{Kirkman2003} and references therein; \citealp{PlanckCollaboration2016}). This is consistent with the number of baryons observed during the first billions of years of the Universe in the Ly$\alpha$ Forest (e.g., \citealp{Michael1998}). However, in the local Universe ($z<2$), the number of baryons detected is significantly smaller than predictions (e.g., \citealp{Shull2012}). At galactic scales, it has been observed that for galaxies with luminosity less or near to Schechter L$^{\star}$, there is a lack of mass of around 50$\%$. For less massive galaxies, larger discrepancies are found (e.g., \citealp{McGaugh2010}).

    Coupled with the missing baryons problem, at galactic scales, we also find the missing metal problem. If we calculate the number of reprocessed metals estimated by the number of stars observed and the star formation history of the Milky Way (MW) disk, we find that the resulting value would be above the number of detected metals we can observe. For example, \citet{Peeples2014} found that for galaxies with a stellar mass around 10$^{10}$ M$_\odot$, only 40$\%$ of the metals produced in stars can be detected up to distances of tens of kpc. 

    Theoretical studies suggest that the missing baryons and missing metals in galaxies might reside in the CGM. However, this medium is very hot and diffuse (e.g., \citealp{Feldmann2013}), therefore it is not easy to detect, especially in emission. Hence, a common way of studying it is by viewing it in absorption against bright background sources such as quasars or blazars \citep{Tumlinson2017}. This method allows the detection of faint signals from absorption lines on their spectra (e.g., \citealp{Mathur2022}).

    \cite{Gupta2012} investigated the absorption lines of O $\mathrm{VII}$ and O $\mathrm{VIII}$ in the CGM of our Galaxy, by measuring the column densities from the K$\alpha$ and K$\beta$ transitions to derive properties of the hot gas around the Galaxy. They found the presence of \warmhot\ gas at temperatures of 10$^6$ K permeating the region above and below the Galactic disk up to distances of 100 kpcs. The total mass comprised in the CGM is above 10$^{10}$ M$_{\odot}$, an order of magnitude greater than previously thought. They conclude that there is an extensive reservoir of ionized gas in the halo around the Milky Way, possibly solving the missing baryon problem at least in our own Galaxy.
    
    However, recent evidence has shown that the Galactic halo is more complex than previously thought. In addition to the \warm\ and \warmhot\ components, the presence of a third gaseous component at a super-virial temperature in the Milky Way CGM has been recently discovered by \citet{Das2019} \& \citet{Das2021}. They found in absorption the super-virial phase along two lines of sight towards individual quasars. Subsequently, the super-virial \hot\ component of the MW CGM was also detected in emission (\citealp{Gupta2020}; \citealp{Bhattacharyya2022}; \citealp{Kaaret2019}). This \hot\ component, at $T \sim 10^7$ K, is not expected from theoretical studies which predict only the presence of gas at or below the virial temperature. 
    
    In a very comprehensive study, \citet{Das2021} detected several ionizing species of this super-virial \hot\ phase of the Milky Way CGM using Chandra X-Ray Observatory (\Chandra) data toward Mrk 421. They found traces of \NeX\ \Ka\ ($\lambda$ 12.132 \AA) and \SiXIV\ ($\lambda$ 6.182 \AA), which relate to this novel \hot\ component. These results suggest that the missing baryons and missing metals could reside not only in the \warm\ and \warmhot\ phases, but also in this component. However, the nature of this gas phase and the total amount of gas in it are still far from being understood.
    
    In this work, we explore this \hot\ component of the CGM by using a stacking technique over multiple grating X-ray spectral observations towards different lines of sight. We focus our study in the absorption lines of highly ionized elements within the 4 to 8 \AA\ wavelength range. As noted in $\S$~\ref{sec:dataSample}, we have removed sources with high signal-to-noise ratio (S/N) spectra from our sample; our focus here is on the power of the stacking technique, using low S/N in individual spectra in which absorption lines could not be detected.

    This paper is structured as follows: in $\S$~\ref{sec:dataSample} we present the sample selection. In $\S$~\ref{sec:analysis} we describe the data analysis. In $\S$~\ref{sec:results} we present our results, and in $\S$~\ref{sec:discussion} we discuss their implications.
    
%
%
    
\section{Data Sample \label{sec:dataSample}}

    Our sample consists of all X-ray grating public observations of AGNs and Quasars available up to June 11, 2022 that are found in both, \href{https://cda.harvard.edu/chaser/}{\Chandra} and \href{https://bit.ly/3I7A3d8}{\XMMN} databases. 

    To study the hot CGM, we focus only on the spectral range between $\sim$ 4 - 8 \AA, where the ionic transitions of \SiXIV\ \Ka\  ($\lambda$ 6.182 \AA) and \SXVI\ \Ka\ ($\lambda$ 4.729 \AA) are found. Because of this, in the current work, we present the analysis of the \ACISSHETGMEG\ (hereafter \MEG)  and \ACISSLETG\ (hereafter \LETG) data, which are the only instruments with good throughput below 12 \AA.\footnote{The analysis of all the components of the CGM, using the complete \XMMN\ and \Chandra\ sample will be presented in Lara-DI et al. in preparations. }  

    We found 47 targets with \MEG\ and/or \LETG\ data. For \MEG\ we used observations through 46 different sight lines with a total exposure time of 10.96 Ms. For \LETG\ we have nine observed targets (1.09 Ms).

    Table~\ref{tab:targets} lists our final sample. In the first column, we name the target. The second column is the redshift of the object. The third and fourth columns show in decimal units their right ascension (RA) and declination (DEC) Equatorial coordinates (ep=J2000), respectively. The Aitoff projection of these targets is shown in Figure~\ref{fig:AitoffMap}. The complete table with the list of individual observations can be found in the Appendix~\ref{app:obsids}.

    This sample only includes QSOs, Seyfert-1, and Blazars. Our sample does not include NGC 4051 with intrinsic absorptionfalling at the position of the z = 0 systems. It also excludestargets with large S/N that would dominate the stacked spectrain our analysis and then affect our results (Mkn 421, 3C 273, and PKS 2155-304).

\begin{table}[ht!]
    \centering
    \renewcommand{\arraystretch}{0.75}\addtolength{\tabcolsep}{2pt}
    \caption{Data Sample. \label{tab:targets}}
    \begin{tabular}{lccc}
        \hline
        {Target}  & {z} & {RA} & {DEC} \\ \hline \hline
    1ES 1028+511	&	0.360	&	157.827	&	50.893	\\
1H0414+009	&	0.287	&	64.218	&	1.090	\\
1H0707-495	&	0.041	&	107.173	&	-49.552	\\
1H1426+428	&	0.129	&	217.136	&	42.672	\\
3C 111	&	0.050	&	64.589	&	38.027	\\
3C 120	&	0.034	&	68.296	&	5.354	\\
3C 279	&	0.536	&	194.046	&	-5.789	\\
3C 382	&	0.058	&	278.764	&	32.696	\\
3C 390.3	&	0.056	&	280.538	&	79.771	\\
3C 445	&	0.056	&	335.956	&	-2.104	\\
3C 454.3	&	0.859	&	343.490	&	16.148	\\
4C 74.26	&	0.104	&	310.655	&	75.134	\\
Ark 120	&	0.033	&	79.047	&	-0.150	\\
Ark 564	&	0.025	&	340.664	&	29.725	\\
ESO 198-G24	&	0.045	&	39.582	&	-52.193	\\
Fairall 51	&	0.014	&	281.225	&	-62.365	\\
Fairall 9	&	0.047	&	20.942	&	-58.806	\\
H1821+643	&	0.297	&	275.489	&	64.343	\\
IC 4329A	&	0.016	&	207.330	&	-30.309	\\
IRAS 13349+2438	&	0.108	&	204.328	&	24.384	\\
MCG-6-30-15	&	0.008	&	203.974	&	-34.296	\\
MCG8-11-11	&	0.021	&	88.723	&	46.439	\\
MR 2251-178	&	0.064	&	343.524	&	-17.582	\\
Mkn 1040	&	0.017	&	37.060	&	31.312	\\
Mkn 279	&	0.031	&	208.264	&	69.308	\\
Mkn 290	&	0.030	&	233.968	&	57.902	\\
Mkn 335	&	0.025	&	1.581	&	20.203	\\
Mkn 509	&	0.034	&	311.041	&	-10.723	\\
Mkn 705	&	0.029	&	141.514	&	12.734	\\
Mkn 766	&	0.013	&	184.610	&	29.813	\\
Mkn 841	&	0.036	&	226.005	&	10.438	\\
Mkn 876	&	0.121	&	243.488	&	65.719	\\
NGC 3227	&	0.004	&	155.877	&	19.865	\\
NGC 3516	&	0.009	&	166.697	&	72.569	\\
NGC 3783	&	0.010	&	174.757	&	-37.739	\\
NGC 4151	&	0.003	&	182.636	&	39.405	\\
NGC 5548	&	0.016	&	214.498	&	25.137	\\
NGC 7469	&	0.017	&	345.815	&	8.874	\\
NGC 985	&	0.043	&	38.657	&	-8.787	\\
PDS 456	&	0.184	&	262.082	&	-14.266	\\
PG 0844+349	&	0.064	&	131.927	&	34.751	\\
PG 1211+143	&	0.081	&	183.573	&	14.053	\\
PG 1404+226	&	0.098	&	211.591	&	22.396	\\
PKS 1830-211	&	2.507	&	278.416	&	-21.061	\\
PKS 2149-306	&	2.345	&	327.981	&	-30.465	\\
Q0836+7104	&	2.172	&	130.351	&	70.895	\\
TON S 180	&	0.062	&	14.334	&	-22.382	\\
        \hline
    \end{tabular}
\end{table}
    
\begin{figure*}
    \includegraphics[width=\textwidth]{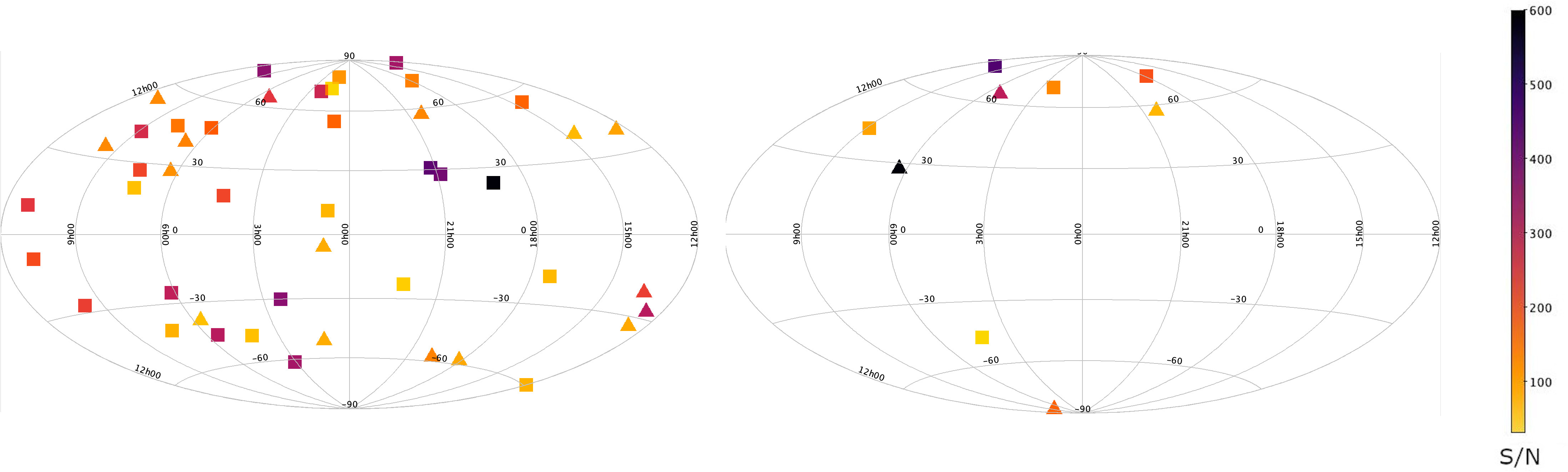}
    \caption{Aitoff Projection of \MEG\ data (left) and \LETG\ data (right) targets in galactic coordinates. Targets with warm absorbers are shown with squares. Targets without warm absorbers are shown in triangles. The S/N ratio is in color-bar code and was calculated in the wavelength range from 4 to 8 \AA}
    \label{fig:AitoffMap}
\end{figure*}

    \subsection{Targets Data Reduction}

        We downloaded and reprocessed the data following the \Chandra\ reduction threads and using the \chandrarepro\ script with the \Chandra\ Interactive Analysis of Observations (\CIAO) software (v4.13) and \Chandra\ Calibration Database (\CALDB, v4.9.6). This script ``reads data from the standard data distribution and creates a new bad pixel file, a new level=2 event file, and a new level=2 Type II PHA file with the appropriate response file'', as mentioned in \href{https://cxc.cfa.harvard.edu/ciao/ahelp/chandra_repro.html}{AHELP for CIAO 4.14}.

%
%
        
\section{Analysis \label{sec:analysis}}

    %
    %

    \subsection{Stacking \label{subsec:stacking}}
        The CGM at $T>10^7$ K is extremely diffuse, making the absorption lines very shallow and faint. This makes their detection very challenging. To overcome this limitation we used the stacking technique over our \MEG\ and \LETG\ samples. This technique improves the S/N of the stacked spectrum by adding counts from individual observations, making the detection of faint signals possible. 

        Using the \combinegratingspectra\ command on the \CIAO\ software, we stacked the observations of each sub-sample listed in the Appendix~\ref{app:obsids}, using the 1st grating order spectra. The \combinegratingspectra\ command combines \Chandra\ gratings PHA files and produces a weighted average of their associated response files (ARF and RMF). It also combines the associated background PHA spectra and source and background ARF and RMF files. 

        In the present work, we are focusing in the detection of the ionic transitions of \SiXIV, and \SXVI, which probe the \hot\ component of the Milky Way. The S/N per resolution element (SNRE) around 0.1 \AA\ of \SiXIV\ \Ka\ is 373 for \MEG\ and 103 for \LETG.
           
    %
    %

    \subsection{Line modeling \label{subsec:fittingAbsorptionLines}}

        We performed the spectral fitting using the software \XSPEC\ (v12.12.0) and chi-squared statistics, focusing in the absorption lines of \SiXIV\ \Ka\ and \SXVI\ \Ka, associated to the \hot\ component of the CGM at $T>10^{7}$ K. Errors were calculated within a confidence range of 1$\sigma$.

        We constrained our modelling to the local spectrum in the range around $\pm$ 0.25 \AA\ from the position of the rest-wavelength of each ionic transitions of interest. First, we fit the local continuum in this range using a power-law (\XSPEC\ model \pow). Then, we included as many Gaussians as were necessary to fit intrinsic AGN absorption or emission features.
    
        Once we had a good model description of the local continuum plus AGN lines, we focused on modeling the lines from the \hot\ CGM. To account for the $z=0$ absorption, we added a narrow Gaussian profile at the rest-frame wavelength of each of the absorption lines from the CGM ions. We set the position of each line to its expected value, and let it vary around the size of the Resolution Element\footnote{Except for \SiXIV\ \Ka\ in \MEG\ data. In this case given that there are residuals around the line, we constrain the position of the line to vary within 0.004 \AA.}. We fixed the width of the line to 0 \AA, and let the normalization free to vary. Finally, using the \steppar\ command, an inbuilt \XSPEC\ function that allows performing a fit while stepping the value of a parameter through a given range, we determined confidence ranges for the best fit of the line and determined its position.
        
    %
    %

    \subsection{Evaluating The Significance of the Line} \label{sec:evaluatingsignifLine}
       
        To quantify the goodenes of the modeled line, we used two different methods. The first method was to estimate its significance ($\sigma$) by dividing its EW by its error on the lower side.
   
        The second method was the inverse of the Akaike Information Criterion (IAIC) \citep{Akaike1998}. This is a method that gives information of how well a model fits the data. It is given by
  
        \begin{equation}
            IAIC = {\exp\left(\frac{(2k_1+\chi_1^2)-(2k_2+\chi_2^2)}{2}\right)}^{-1},
        \end{equation}
  
        where $k_1$ is the number of free parameters that includes the model fitting the continuum plus the absorption line, $\chi^2_1$ is its best fit statistic. $k_2$ is the number of parameters of the model without the line at $z=0$, and $\chi^2_2$ its statistic. 

        By calculating the IAIC we obtain a factor by which the model fitting the local continuum (plus additional AGN lines) and the ionic transition at $z=0$ is preferred over the simple model not including this last ion.   

%
%
        
\section{Results} \label{sec:results} 

\begin{figure}[htpb!]
    \centering
    \renewcommand{\arraystretch}{2}\addtolength{\tabcolsep}{2pt}
    \includegraphics[width=0.415\textwidth]{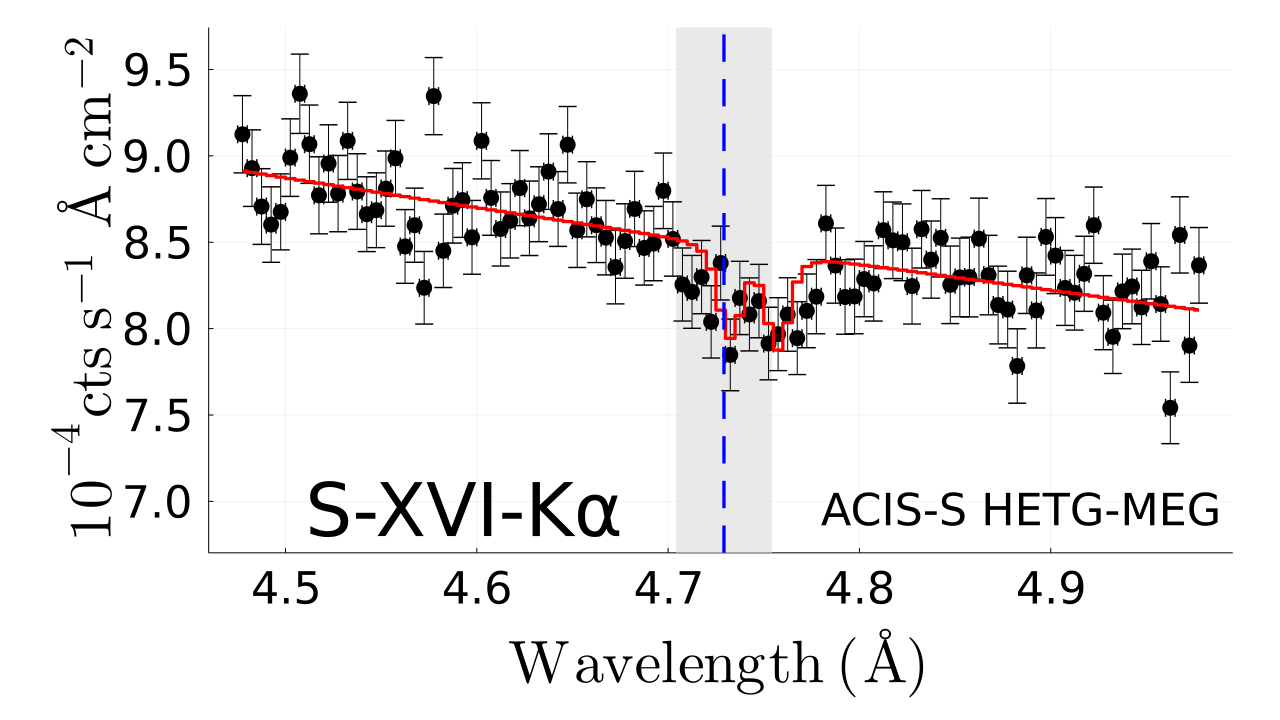}
    \includegraphics[width=0.415\textwidth]{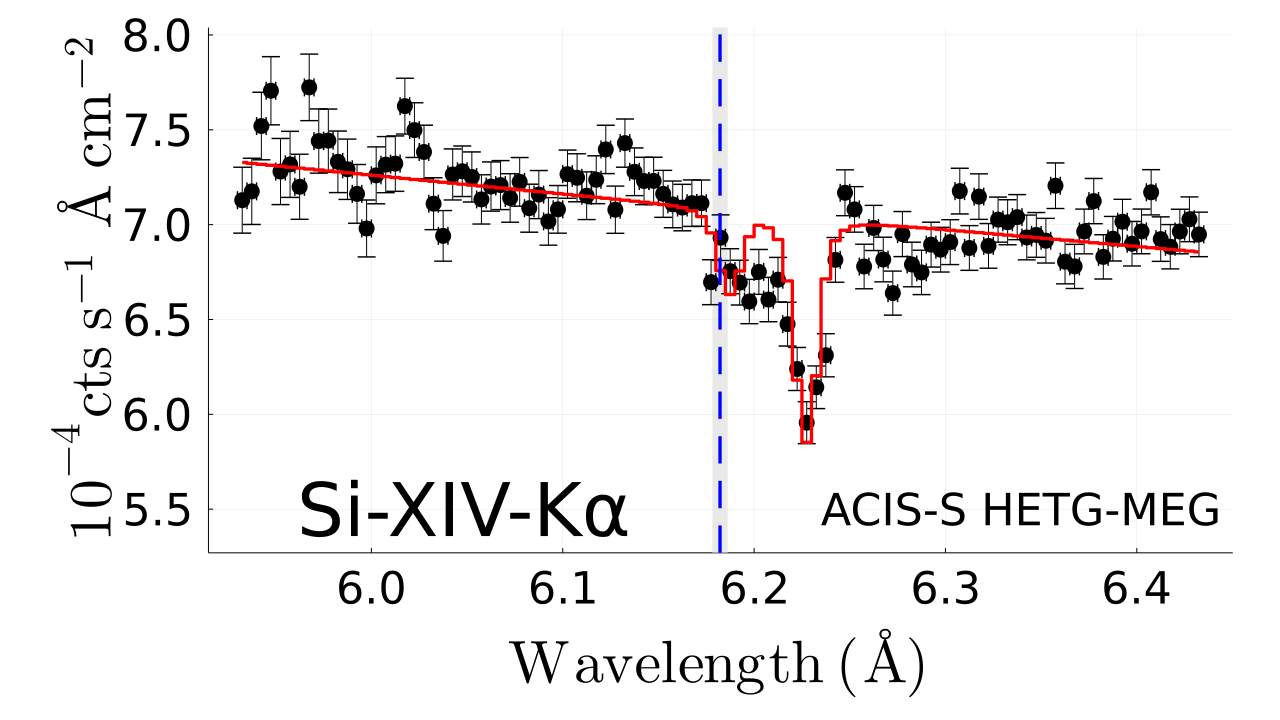}
    \includegraphics[width=0.415\textwidth]{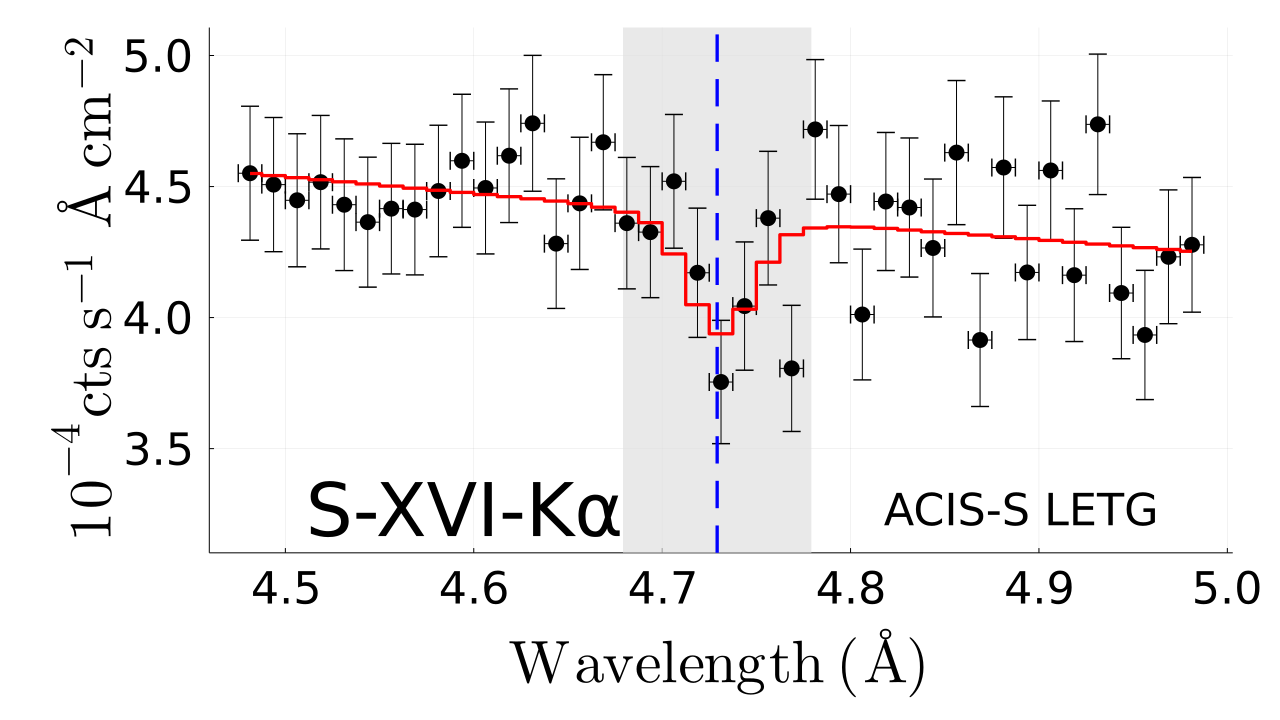}
    \includegraphics[width=0.415\textwidth]{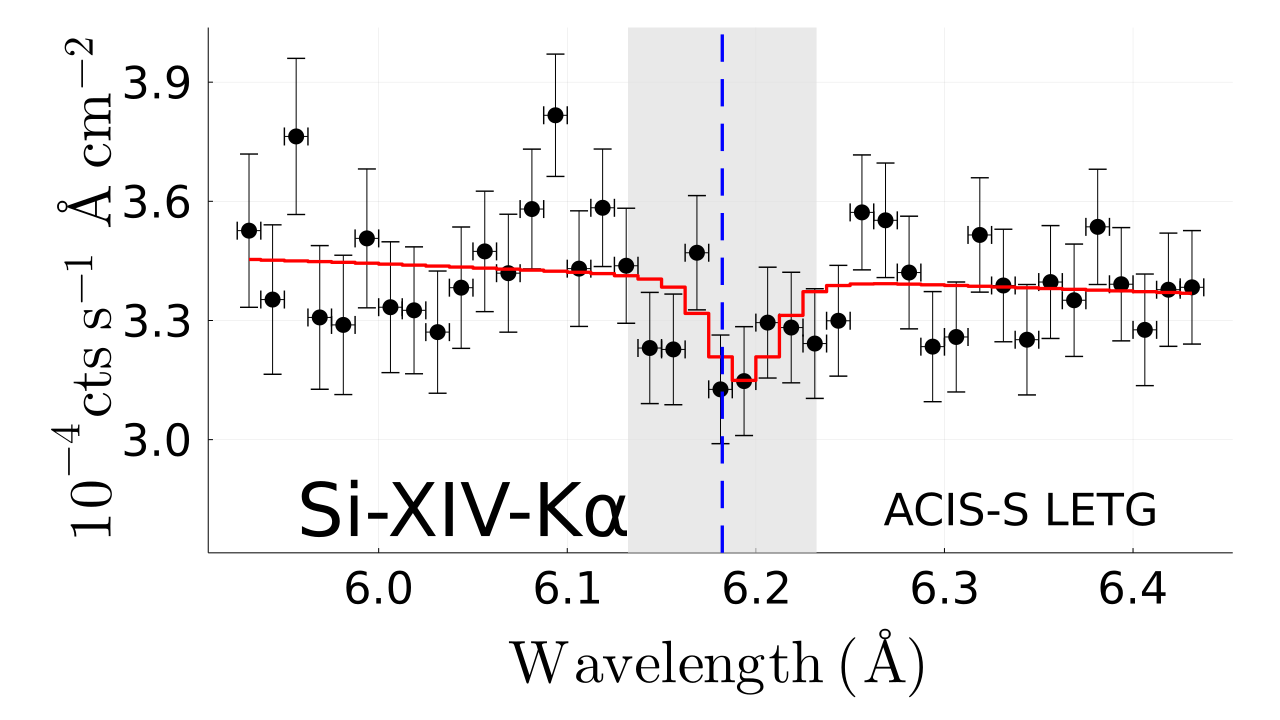}
    \caption{Data is plotted in black, the model in red, and the position of the theoretical rest wavelength with a vertical blue dashed line. The gray area is the the size of the the resolution element of the telescope instrument.}
    \label{fig:results}
\end{figure}

   We detected highly ionized absorption lines at $z=0$ in the stacked spectra along different lines of sight. Our results are presented in Figure~\ref{fig:results}. In \MEG\ we detected \SXVI\ \Ka\ ($3.9\sigma$) and \SiXIV\ \Ka\ ($5.4\sigma$). In \LETG\ we detected \SXVI\ \Ka\ at $2.8\sigma$ and Si XIV \Ka\ at $2.7\sigma$.
   
   The intrinsic WA lines can be seen in Figure~\ref{fig:results}. For example, the absorption line at 6.23 \AA\ in the \MEG\ panel of Figure~\ref{fig:results} is the intrinsic \SiXIV\ \Ka\ line at $z=0.0076$. Identification of the intrinsic lines, and the discussion of the associated WA is deferred to a follow-up paper. For the purpose of this paper, we note that the intrinsic lines are clearly separated from the $z=0$ lines from the MW CGM; there is no ambiguity about the identification of the $z=0$ lines (see \S \ref{sec:discussion}).

   We list our results in Table~\ref{tab:results}, where we show (1) ionic transition, (2) the rest wavelength in \AA, (3) the oscillator strength, (4) The S/N in the $\pm 0.25$ \AA\ range around the theoretical wavelength of the line, (5) the observed wavelength in \AA, (6) the $\chi^2$/d.o.f. statistics of the fit when the model only considers the local continuum and WA intrinsic features (thereafter model c), (7) the $\chi^2$/d.o.f. statistics when the $z=0$ absorption lines are included (thereafter model c+l), (8) the EW (m\AA) and errors of the $z=0$ line, (9) the ionic column density calculated using the curve of growth \footnote{The linear part of the curve of growth consists in the proportional relation between the equivalent width of the line and the ionic column density. For optically thin gas $N(ion) = 1.3\times10^{20}\left(\frac{EW}{f \lambda^2}\right)$, where $N(ion)$ is the ionic column density in ($cm^{-2}$), $EW$ is the equivalent width of the line in \A, $f$ is the oscillator strength of the transition, and $\lambda$ the position of the line at rest-wavelength in \AA.}, (10) the statistical significance of the line ($\sigma$), and (11) the IAIC value.

\begin{deluxetable*}{lcccccccccc}
   \centering 
   \renewcommand{\arraystretch}{1}\addtolength{\tabcolsep}{-1pt}
   \tablecaption{Identified ionic absorption lines probing the super-virial hot component at z=0. \label{tab:results}}{.}
   \tablehead{
      \colhead{Ionic Transition}       & 
      \colhead{$\lambda$}        & 
      \colhead{Oscilator Strength}     & 
      \colhead{S/N}                    &
      \colhead{$\lambda_o$}      &
      \colhead{$\chi_{c}^2$/d.o.f.}    &
      \colhead{$\chi^2_{c+l}$/d.o.f.}  &
      \colhead{EW}              &
      \colhead{N ion}      & 
      \colhead{$\sigma$}               &
      \colhead{IAIC} \\
      \colhead{}       & 
      \colhead{[\AA]}        & 
      \colhead{}     & 
      \colhead{}                    &
      \colhead{[\AA]}      &
      \colhead{}    &
      \colhead{}  &
      \colhead{[m\AA]}              &
      \colhead{[$10^{16}$cm$^{-2}$]}      & 
      \colhead{}               &
      \colhead{} 
   }
   \colnumbers
   \startdata
      \hline \hline
      \multicolumn{11}{l}{{\fontfamily{qcr}\selectfont Instrument:{ACIS-S}, Grating:{HETG-MEG}, Exposure Time:{1.09E+07sec}, Resolution Element:{25m\AA}}} \\ \hline	
      S-XVI-K$\alpha$ & 4.729 & 0.416 & 89.88 & 4.735 & 110.49/99 & 97.0/96 & 1.07$^{+0.32}_{-0.27}$ & 1.50$^{+0.44}_{-0.38}$ & 3.9 & 313  \\
      Si-XIV-K$\alpha$ & 6.182 & 0.416 & 127.07 & 6.186 & 196.14/97 & 167.26/96 & 1.06$\pm 0.20$ & 0.87$\pm{0.16}$ & 5.4 & 6.87$\times 10^{5}$  \\
      \hline \hline
      \multicolumn{11}{l}{{\fontfamily{qcr}\selectfont Instrument:{ACIS-S}, Grating:{LETG}, Exposure Time:{1.09E+06sec}, Resolution Element:{50m\AA}}} \\ \hline	
      S-XVI-K$\alpha$ & 4.729 & 0.416 & 36.99 & 4.732 & 34.43/39 & 26.85/36 & 4.87$^{+1.79}_{-1.75}$ & 6.80$^{+2.51}_{-2.44}$ & 2.8 & 16.3  \\
      Si-XIV-K$\alpha$ & 6.182 & 0.416 & 48.62 & 6.199 & 37.38/39 & 30.05/36 & 3.49$^{+1.28}_{-1.30}$ & 2.85$^{+1.05}_{-1.06}$ & 2.7 & 14.4  \\
   \enddata
 \end{deluxetable*}         

%
%
   
\section{Discussion \label{sec:discussion}}

    %
    %

    \subsection{Detection of the Hot Component of the CGM \label{subsec:discussion_detectionHotComponent}}
    
        The high S/N of the data, particularly in \MEG, allowed us to detect two different transitions at high level of significance ($>3\sigma$) from the \hot\ component of the CGM. Namely, the transitions by \SXVI\ \Ka\ (3.9$\sigma$), and the \SiXIV\ \Ka\ (5.4$\sigma$). These two \Ka\ transitions are further confirmed, but with lower confidence, in \LETG. The significance is 2.8$\sigma$ for \SXVI\ and 2.7$\sigma$ for \SiXIV. The lines were detected with much higher significance with \MEG\ than with \LETG\ because of the total observing time available for the stacking with each detector. \MEG\ data has one order of magnitude larger exposure, which translates in a factor of 3.8 in S/N. Also, \LETG\ has a factor of 2 larger resolution element than \MEG, making the EW sensitivity worse. This leads to the difference in the detection significance of lines, as expected.   
        
        The formal level of significance including both \ACISS\ \LETG\ and \MEG\ data (adding in quadrature individual significances) is 4.8$\sigma$ for \SXVI\ \Ka, and 6.0$\sigma$ for the \SiXIV\ \Ka. Thus, the presence of a \hot\ CGM component is detected with high level of confidence. 
        
        To further evaluate this detection, we note that our sample includes objects with intrinsic absorption produced by WA winds in their individual spectra. Thus, it is mandatory to show that the detected CGM lines arise from our Galaxy and not from these outflows. This is particularly important for \MEG, since 90$\%$ of the observing time (S/N$\sim$ 1345) comes from objects with detected intrinsic absorbers. As can be seen in Figure~\ref{fig:results} for the \MEG\ data, the sum of the intrinsic absorption in the stacked spectrum is clearly seen in the form of broad absorption lines redwards of the $z=0$ position. These intrinsic absorption features are located at an average redshift of $\sim 0.04 - 0.05$ \AA, or $> 2,000$ km s$^{-1}$, with respect to our rest-frame. Thus, these lines are clearly not blended with the CGM absorption. For \LETG\ the situation is different. The observing time is dominated by objects without WA (observing time 56$\%$ and S/N$\sim$ 292). This, in addition to the limited S/N for this detector, is the reason that intrinsic WA absorption is not detected in this stacked spectrum. Finally, we emphasize that there are no WA in the individual objects whose outflow velocity might be coincident with our unique position at $z=0$. The only source where WA absorption lines would fall at the position of the $z=0$ lines from the MW CGM is NGC 4051, and these data were explicitly excluded from our analysis as discussed in $\S$~\ref{sec:dataSample}. In addition, all the $z=0$ lines detected are narrow. This gives us great confidence that the ionic transitions at $z=0$ arise from the hot component of the CGM of our own Galaxy.

    %
    %

    \subsection{The Extended \hot\ CGM Component \label{subsec:discussion_extendedHotComponent}}

	    The hot CGM was detected by \citealp{Das2019} and \citealp{Das2021} in the sightlines towards blazars 1ES 1553+113 and Mkn 421, respectively. They associated this plasma with the \hot\ component of the a CGM at $T \approx 10^{7.5}$ K. These blazars are among the most luminous in X-rays and their \Chandra\ data would have had a large weight in the S/N of our observations, because of very extended observing time and/or because the data was taken during  extreme bursts (Mrk 421). For this reason, we explicitly excluded these objects from our sample as discussed in $\S$~\ref{sec:dataSample}. Therefore, the results presented here are new and independent detections to that present in the literature. 

        The high-significance detection of \SiXIV\ in our data confirms the presence of the \hot\ component of the Milky Way CGM. The \SXVI\ \Ka\ is the first detection of the \hot\ component of the CGM to date. This line could come from the same component as \SiXIV\ \Ka\ or even a hotter phase. The geometry of the \warmhot\ component of the CGM is that of a uniform gas extending through vast regions of the galactic halo. It has been shown that the covering factor of this component is at least 72$\%$ (\citealp{Gupta2012}, \citealp{Mathur2012}). The \warm\ component, although patchy, is also extended throughout the CGM, with a covering factor of at lest 80$\%$ (\citealp{Tumlinson2011}). The detection of the \hot\ component in the only two blazars with enough sensitivity two see this phase, and in the stacked average spectra of few tens of objects towards different lines of sight, points to the \hot\ gas spreading throughout the halo.  This indicates that the covering factor of the super-virial temperature CGM must be significant.
        
        Our results further confirm that the hot CGM is not only related to the physical processes forming the Galactic bubbles (\citealp{Predehl2020}; \citealp{Su2010}). See Figure~\ref{fig:AitoffMap}. Our averaged spectra include an ample set of lines of sight, including many directions that do not cross these bubbles.  The same was found by \citealp{Das2021} for Mrk 421, whose line of sight does not cross the bubbles. This is further confirmed by emission studies of the CGM, that have found a hot component coexistent with the virial phase towards directions away from the Galactic Center (e.g. \citealp{Gupta2022}; \citealp{Bhattacharyya2022}; \citealp{Bluem2022}).

    %
    %

    \subsection{Spatial Distribution of the Hot Phase of the CGM\label{subsec:discussion_spatialDistribution}}
        
        Using the curve of growth, we calculated the ionic column density of \SiXIV\ \Ka\ (N$_{Si14}$) and \SXVI\ \Ka\ (N$_{S16}$). Table~\ref{tab:results} shows that \LETG's best-fit N$_{S16}$ is $\approx$ 4.5 greater than \MEG's. The same occurs with the best-fit value of N$_{Si14}$, being $\approx$ 3.3 times grater in \LETG\ than in \MEG. We explore different possibilities for these differences.
        
        First, our \MEG\ and \LETG\ spectra corresponds to two different samples. \MEG\ sample corresponds to a stacked spectra of 46 targets, while \LETG\ only to nine (see Table~\ref{tab:ObsIDs}). Among these nine \LETG\ targets, only one is not included in \MEG\ sample: TON S 180  corresponding to 7$\%$ of the total \LETG\ exposure time. Thus, giving the different composition of the \MEG\ and  \LETG, we might expect that the difference in the column density is due to the different lines of sight averaged in these samples.
       
        However, we note that \MEG\ have larger exposure times ($\approx$ 10) and S/N ($\approx$ 3.8) than \LETG. This results in much larger errors in the latter sample. If we compare the column densities obtained with \LETG,  they are consistent within $\approx$ 2$\sigma$ with the \MEG\ values. This may be indicating that the difference between \MEG\ and \LETG\ arises from the low significance and large errors in the \LETG\ data.
        
        In the line of sight towards Mkn 421, \citet{Das2021} detected \SiXIV\ \Ka, and reported a column density of $7.48 \pm 2.08 \times 10^{15} \mathrm{cm}^{-2}$. This value is strikingly similar to the average value measured over the \MEG\ sample. This suggests that the \hot\ CGM detected in absorption may be as homogeneous as the warmer components. However, our analysis is too limited to further explore this possibility. 
        This result would be in contrast to the Emission Measure of the CGM \hot\ and warm phases which have been observed to vary by an order of magnitude in large and short scales  (\citealp{Das2019}, \citealp{Gupta2021}, \citealp{Gupta2022}; \citealp{Bluem2022}; \citealp{Bhattacharyya2022}).
        
        Studying in detail the gas distribution of the \hot\ CGM phase requires much better data with significant detections among individual lines of sight. 
        
        Finally, we emphasize that in this work we are reporting the first detection of \SXVI. 
        
%
%

\section{Conclusion} \label{sec:conclusions} 

    The detection of highly ionized absorption lines through the stacking of multiple sightlines confirms the presence of a novel \hot\ component of the Milky Way CGM. This \hot\ phase extends over a significant fraction of the CGM and was not detected by  chance in a few lines of sight. The overall detection of different ions for this phase points to the possibility that the \hot\ gas component of the Milky Way may contain a significant fraction of the missing baryons and missing metals at galactic scales. 

%
%

\section{Acknowledgments} \label{sec:acknowledgments} 

    S.M. is grateful for the grant provided by the National Aeronautics and Space Administration through Chandra Award Number AR0-21016X issued by the Chandra X-ray Center, which is operated by the Smithsonian Astrophysical Observatory for and on behalf of the National Aeronautics Space Administration under contract NAS8-03060. S.M. is also grateful for the NASA ADAP grant 80NSSC22K1121

    S.D. acknowledges support from the Presidential Graduate Fellowship from the Ohio State University, and Kavli Fellowship from KIPAC, Stanford University. 
        
%
%

\bibliography{PhD01ScientificArticle01v02}{}
\bibliographystyle{aasjournal}

%
%

\appendix

    %
    %

    \section{List of ObsIDs\label{app:obsids}}
    
    \begin{deluxetable*}{lll}[htp!]
        \center
        \renewcommand{\arraystretch}{0.5}\addtolength{\tabcolsep}{-2pt}
        \tablewidth{0pt}
        \tablecaption{ObsIDs \label{tab:ObsIDs}}{.}
        \tablehead{
                    \colhead{Target}  & 
                    \colhead{Exposure Time (ks)} &
                    \colhead{ObsIDs} 
            }
        \colnumbers \
        \startdata
\hline \hline
\multicolumn{3}{c}{{\fontfamily{qcr}\selectfont {ACIS-S HETG-MEG (10.96 Ms)}}}  \\ \hline	
    1ES 1028+511  & 89.93 & 2970	3472													 \\
1H0414+009  & 86.53 & 2969	4284													 \\
1H0707-495  & 152.35 & 2304	12115	12116	12117	12118										 \\
1H1426+428  & 139.75 & 3568	6088													 \\
3C 111  & 143.41 & 16219														 \\
3C 120  & 251.86 & 3015	16221	17564	17565	17576										 \\
3C 279  & 110.66 & 2971	7360	7361												 \\
3C 382  & 118.02 & 4910	6151													 \\
3C 390.3  & 147.67 & 16220	16530	16531												 \\
3C 445  & 413.22 & 13305	13306	13307	14194											 \\
3C 454.3  & 4.12 & 7362	7363													 \\
4C 74.26  & 68.65 & 4000	5195													 \\
Ark 120  & 120.5 & 15636	16539	16540												 \\
Ark 564  & 294.55 & 863	9898	9899	10575											 \\
ESO 198-G24  & 149.48 & 4817	5315													 \\
Fairall 51  & 233.69 & 20776	20777	21107	21108	22266										 \\
Fairall 9  & 78.93 & 2088														 \\
H1821+643  & 99.62 & 1599														 \\
IC 4329A  & 233.15 & 2177	19744	20070	20095	20096	20097									 \\
IRAS 13349+2438  & 294.52 & 4819	4820													 \\
MCG-6-30-15  & 521.8 & 4759	4760	4761	4762											 \\
MCG8-11-11  & 118.15 & 12861	13200													 \\
MR 2251-178  & 539.14 & 2977	12828	12829	12830											 \\
Mkn 1040  & 197.88 & 15075	15076	16571	16584											 \\
Mkn 279  & 213.32 & 3062	24662	24665	24666	24667										 \\
Mkn 290  & 247.27 & 3567	4399	4441	4442											 \\
Mkn 335  & 230.06 & 23292	23297	23298	23299	23300	23301	23302								 \\
Mkn 509  & 326.93 & 2087	13864	13865												 \\
Mkn 705  & 20.99 & 4914														 \\
Mkn 766  & 422.52 & 1597	16310	16311	16628	16629	16630	16632								 \\
Mkn 841  & 193.59 & 12165	12166	13258												 \\
Mkn 876  & 238.82 & 18144	18145	18146	18147	18148	18799	18814	18844	19885	19889	20022	20047			 \\
NGC 3227  & 194.61 & 860	18618	18849												 \\
NGC 3516  & 386.19 & 2080	2431	2482	7281	7282	8450	8451	8452							 \\
NGC 3783  & 1202.15 & 373	2090	2091	2092	2093	2094	14991	15626	18192	19694					 \\
NGC 4151  & 679.96 & 335	3052	3480	7829	7830	16089	16090	25025							 \\
NGC 5548  & 401.09 & 837	3046	21694	21846	22207	22681									 \\
NGC 7469  & 383.78 & 2956	3147	18622	18623	18733	18734	18735	18736	18737	18738					 \\
NGC 985  & 76.46 & 3010														 \\
PDS 456  & 142.85 & 4063														 \\
PG 0844+349  & 141.18 & 5599	6244	6245												 \\
PG 1211+143  & 433.33 & 17108	17109	17110	17645	17646	17647									 \\
PG 1404+226  & 79.12 & 812														 \\
PKS 1830-211  & 171.49 & 371	2417	22197	22198	22199	22239	22240								 \\
PKS 2149-306  & 89.63 & 336	1481													 \\
Q0836+7104  & 72.35 & 334	1450	1802												 \\
\hline \hline
\multicolumn{3}{c}{{\fontfamily{qcr}\selectfont {ACIS-S LETG (1.09 Ms)}}}  \\ \hline	
1H1426+428  & 39.84 & 6089														 \\
3C 279  & 28.13 & 6867														 \\
3C 445  & 198.56 & 10405	10406	12000												 \\
H1821+643  & 470.15 & 2186	2310	2311	2418											 \\
NGC 3516  & 43.64 & 831														 \\
NGC 4151  & 83.59 & 3089														 \\
NGC 5548  & 19.44 & 15659	15660	15661	15662											 \\
PG 1211+143  & 133.6 & 4880	5329	5330												 \\
TON S 180  & 76.75 & 811														 \\
\enddata
\end{deluxetable*}

%
%

\end{document}